# de Haas–van Alphen study on three-dimensional topological semimetal pyrite PtBi$_2$

Wenshuai Gao[1,2] [*], Xiangde Zhu[1] [*], Hongwei Zhang[1,2], Min Wu[1,2], Jin Hu[5], Fawei Zheng[6], Guolin Zheng[1,2], Ning Hao[1], Wei Ning[1, §], and Mingliang Tian[1,3,4, §]

[1]*Anhui Province Key Laboratory of Condensed Matter Physics at Extreme Conditions, High Magnetic Field Laboratory, Chinese Academy of Sciences, Hefei 230031, Anhui, China;*

[2]*Department of physics, University of Science and Technology of China. Hefei 230026, China;*

[3]*Department of Physics, School of Physics and Materials Science, Anhui University, Hefei 230601, China*

[4]*Collaborative Innovation Center of Advanced Microstructures, Nanjing University, Nanjing 210093, China*

[5]*Department of Physics, Institute for Nanoscience and Engineering, University of Arkansas, Fayetteville, AR 72701, USA*

[6]*Institute of Applied Physics and Computational Mathematics, Beijing 100088, China*

[*]W. Gao and X. Zhu contributed equally to this work

[§] To whom correspondence should be addressed. E-mails: ningwei@hmfl.ac.cn or tianml@hmfl.ac.cn

We present the systematic de Haas−van Alphen (dHvA) quantum oscillations studies on the recently discovered topological Dirac semimetal pyrite PtBi$_2$ single crystals. Remarkable dHvA oscillations were observed at field as low as 1.5 T. From the analyses of dHvA oscillations, we have extracted high quantum mobility, light effective mass and phase shift factor for Dirac fermions in pyrite PtBi$_2$. From the angular dependence of dHvA oscillations, we have mapped out the topology of the Fermi surface and identified additional oscillation frequencies which were not probed by SdH oscillations.

Three-dimensional (3D) topological semimetals, including Dirac and Weyl semimetals, have recently attracted intense attention in the condensed matter community[1-4]. These materials exhibit linearly dispersed energy bands near the band touching points, forming bulk Dirac or Weyl cones and leading to exotic surface states[5-7]. In Dirac semimetals such as $Cd_3As_2$[8,9] and $Na_3Bi$[10,11], the band crossings at the Dirac points protected by the crystal symmetry are four-fold degenerate. When spin degeneracy is lifted through breaking time-reversal or inversion symmetry, the Dirac semimetal state evolves to a Weyl semimetal state and each Dirac cone splits to a pair of Weyl cones with opposite chirality[12]. The relativistic fermions hosted by Dirac and Weyl cones usually exhibit non-trivial Berry's phase[13,14], light effective electron mass[13] and ultrahigh mobility[15], which has been demonstrated by quantum oscillation studies. Furthermore, the exotic transport phenomena such as negative magnetoresistance (MR) induced by chiral anomaly, extremely large and linear MR have also been discovered in these materials.

The recently discovered pyrite $PtBi_2$ has been proposed to be 3D topological semimetal with unique electronic structure and extreme large, unsaturated MR[16]. In this work, we report the de Haas–van Alphen (dHvA) oscillations on pyrite $PtBi_2$ single crystals. The dHvA effect provides a powerful tool for probing the Fermi surface topology and has been widely applied to many topological semimetals, such as $Cd_3As_2$[17], $TaIrTe_4$[18], Zr*HM* (*H* = Si, Ge; *M*=S, Se, Te)[19-21] and TaP[22]. In $PtBi_2$, we observed prominent dHvA oscillations at field as low as 1.5 T. Signatures of relativistic fermions in pyrite $PtBi_2$ has been observed, including light effective mass and high quantum mobility. From the angular-dependent dHvA oscillations, we have also revealed 3D Fermi surface of pyrite $PtBi_2$ and identified additional oscillation frequencies which were not probed by previous SdH oscillations.

High-quality pyrite $PtBi_2$ single crystals [top left corner of Fig. 1(a), inset] were prepared by flux method. The excellent crystallinity has been demonstrated by the sharp single crystal X-ray diffraction (XRD) peaks as shown in the Fig. 1(b), with the full width at half maximum of the rocking curve reaching as low as 0.05 ° [Fig. 1(b), inset]. The composition and structure of the pyrite $PtBi_2$ have been examined by the Rietveld refinement of the power XRD spectrum, from which we have obtained the lattice parameter $a = b = c =$ 6.704(5) Å and α = β = γ = 90 ° [top right corner of Fig. 1(a), inset] for our pyrite PtBi$_2$ samples, which is consistent with previous reported results[23].

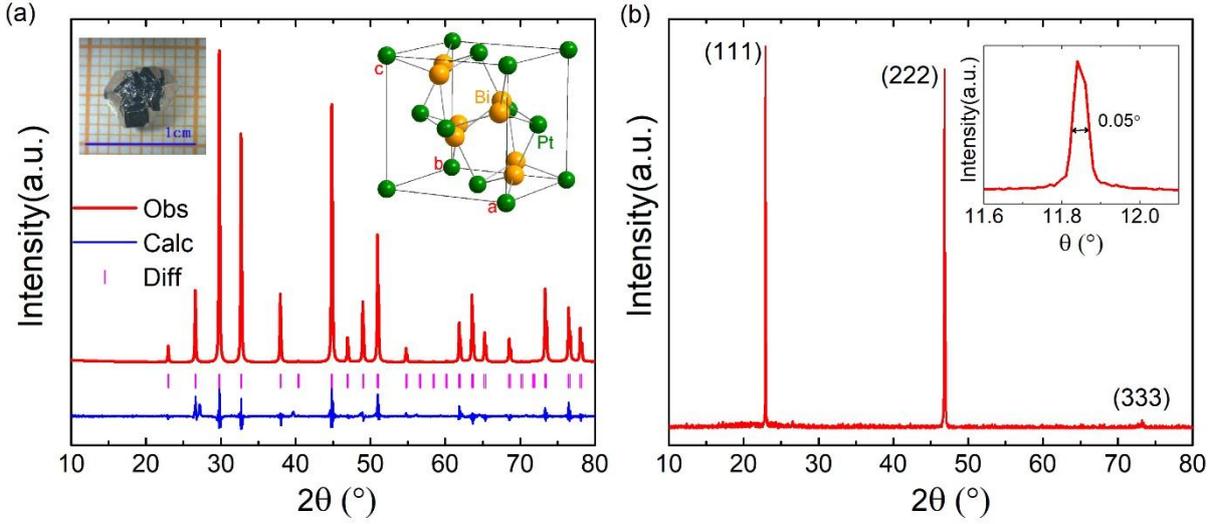

FIG. 1. (a) X-ray diffraction (XRD) pattern of powdered pyrite PtBi$_2$ at room temperature. Inset: image of a single crystal (left) and crystal structure (right). (b) XRD pattern of (111) plane of single crystal sample. Inset: The rocking curve of the (111) diffraction peak, showing the full width at half maximum of 0.05 °.

The dHvA oscillations in pyrite PtBi$_2$ were probed through magnetization measurements by a superconducting quantum interference device (SQUID) magnetometer (Quantum Design). Figure 2(a) shows the magnetization $M$ as a function of magnetic field $B$ when the field is applied perpendicular to (111) plane and $T = 2$ K. Strong magnetization oscillations arise at field as low as 1.5 T [Fig. 2(a)]. The oscillatory component $M$ is striking after removing the smooth magnetization background, as shown in the inset of Fig. 2(b). The fast Fourier transform (FFT) spectrum of $M$ reveals two major frequencies, $F_\alpha = 204T$ and $F_\beta = 730T$. According to the Onsager relation $F = (\hbar/2\pi e)A_F$, the extreme cross section of Fermi surface can be determined to be 0.019(4)Å$^{-2}$ and 0.069(5)Å$^{-2}$ for the α and β bands, respectively. In addition to these two major frequencies, we also noticed two additional frequencies near the 2$^{nd}$ harmonic frequency of $F_\alpha$, marked as $F_{\alpha'}$ and $F_{\alpha''}$ in Fig. 2(b), which has not been observed in the previous SdH oscillation studies[22]. As will be discussed latter, these additional frequencies are result from the hole Fermi pockets in the first Brillouin zone, which enables us to further investigate the electronic properties of pyrite PtBi$_2$.

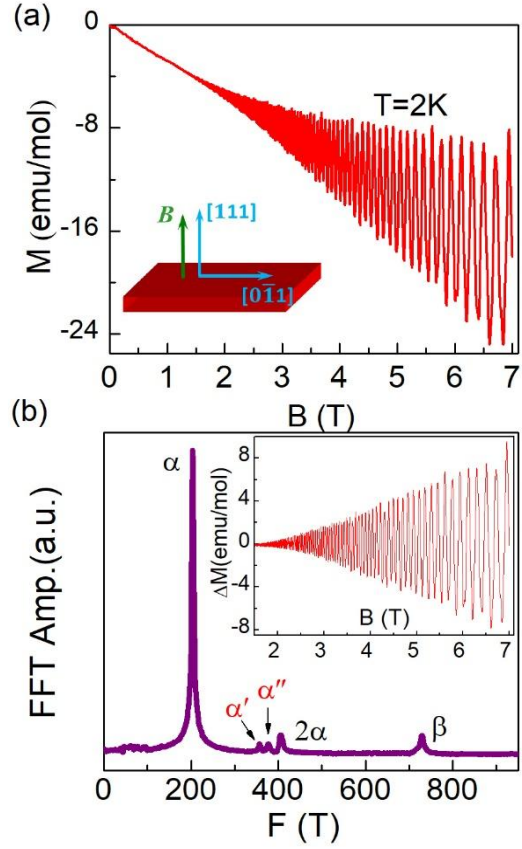

FIG. 2. (a) Magnetic field dependence of the magnetization at $T=2$ K with a magnetic field aligned perpendicular to (111) plane. (b) The corresponding FFT spectrum of the dHvA oscillation. Inset: the oscillatory component of the dHvA effect as a function of $B$.

We have also performed magnetization measurements at various temperatures to obtain the effective mass of the α and β bands. Given the weak F$_β$ oscillation component damps quickly with rising temperature, our measurements are focused on the relative lower temperature range from 1.8 K to 4 K, as shown in Fig. 3(a). The dHvA oscillation that contains phase factor can be described by the Lifshitz-Kosevich (LK) formula[24,25]:

$$\Delta M \propto -B^{1/2} R_T R_D \sin\left[2\pi\left(\frac{F}{B} + \gamma - \delta\right)\right] \quad (1)$$

where $R_T = \alpha T/\sinh \alpha T$, $R_D = \exp(-\alpha T_D m^*/B)$, $T_D$ is the Dingle temperature and $\alpha = 2\pi^2 k_B m^*/\hbar eB$. $k_B$ is the Boltzmann constant, $\hbar$ is the Planck's constant, $m^*$ is the effective cyclotron mass at the Fermi energy, which can be obtained from the fitting of temperature dependence of FFT amplitudes to the thermal damping term of the LK formula, $\frac{2\pi^2 k_B T m^*/\hbar eB}{\sinh[2\pi^2 k_B T m^*/\hbar eB]}$. As shown in Fig. 3(b), our fittings yield the effective masses of $m_\alpha^* = 0.17(8)m_0$ and $m_\beta^* = 0.43(5)m_0$ ($m_0$ is the free electron mass) for the α and β bands, respectively. Both the effective masses are lighter than previous results obtained from the SdH oscillations [22], especially for the α band, whose effective mass is about four times smaller than the transport result.

Besides, by fitting the dHvA oscillation amplitudes with inverse magnetic field to the Dingle damping terms $exp(-2\pi^2 k_B T_D/\hbar\omega_c)$, we can obtain the Dingle temperature of both frequencies. To achieve more accurate fits, we separated the oscillation components of α and β band through filtering out irrelevant oscillations, then we extracted the oscillation amplitude of 2K as a function of 1/B, the best linear fittings based on the transformational LK equation yield the Dingle temperatures $T_D^\alpha = 3.26\pm0.26K$ and $T_D^\beta = 3.87\pm0.31K$, respectively. The quantum scattering lifetime $\tau_Q$, which is related to the Dingle temperature by $\tau_Q = \hbar/2\pi k_B T_D$, are $\tau_Q^\alpha = 3.7(4) \times 10^{-13}s$ and $\tau_Q^\beta = 3.1(5) \times 10^{-13}s$, the quantum mobilities estimated by $\mu_Q = \frac{e\tau_Q}{m^*}$, are $\mu_Q^\alpha = 3940.4\ cm^2V^{-1}s^{-1}$ and $\mu_Q^\beta = 1299.6\ cm^2V^{-1}s^{-1}$, respectively. Such high quantum mobilities are higher than the ones obtained from the SdH oscillations.

In addition to the light effective mass and high mobility, Berry phase close to $\pi$ is another important characteristic for topological non-trivial bands. The Landau level (LL) index fan diagram has been widely used to extract Berry phase for topological materials. However, for oscillations containing multiple frequencies, phase factor for each frequency component is better to be extracted by fitting the oscillation pattern to the LK formula [26]. As shown in the Fig. 3(d) and (e), we have separated the oscillation components of α and β band through filtrating frequencies. With the effective mass, frequency and Dingle temperature of each band extracted from the dHvA oscillations as the fixed parameters, we expected to duplicate the oscillation pattern through adjusting the phase factor. The LK model reproduces each oscillation pattern very well and yields a phase factor $\gamma - \delta$ of 0.038 and -0.61 for α and β band, respectively. Such results are coincident with the transport results that from the SdH oscillations.

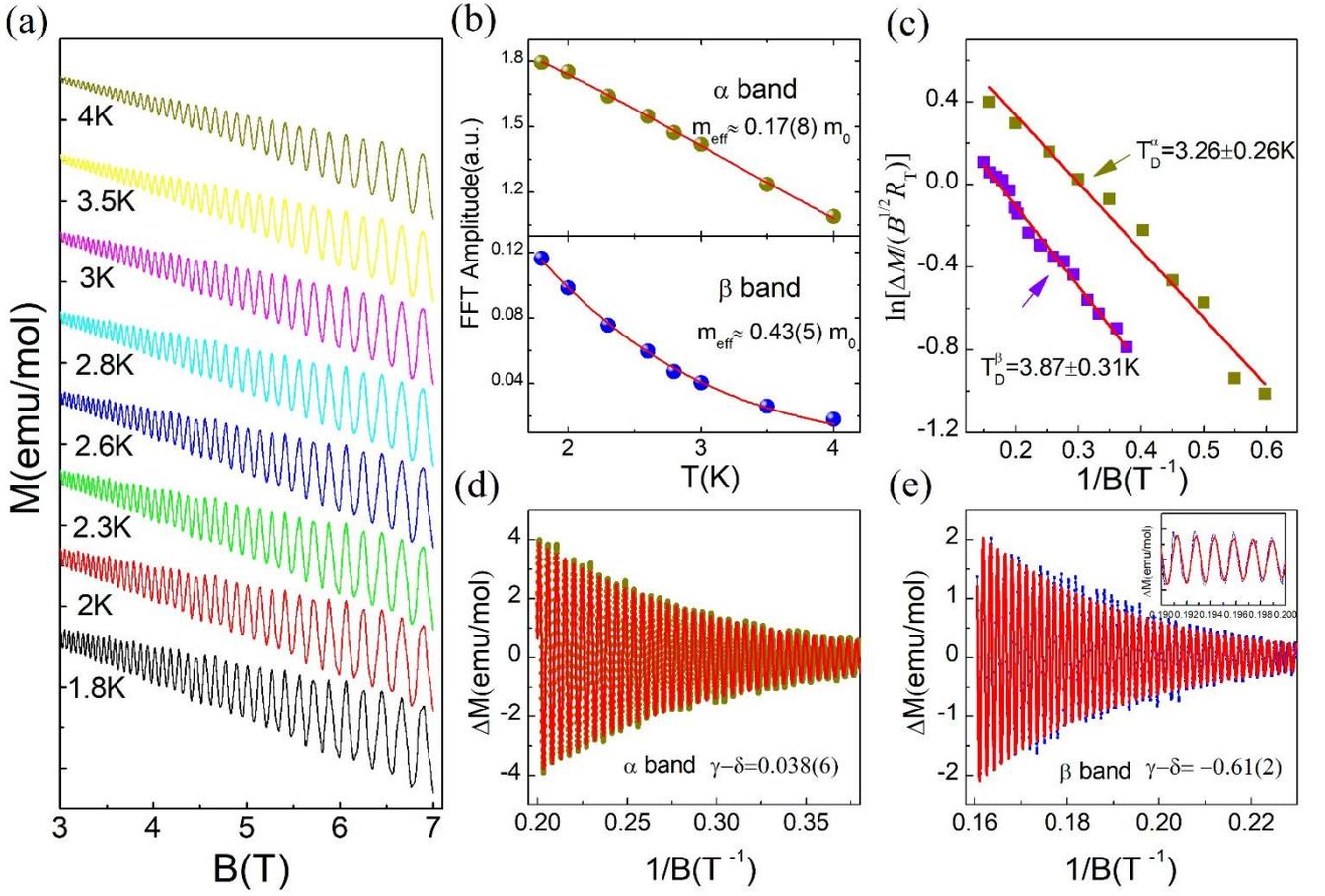

FIG. 3. (a) Magnetic field dependence of magnetization at different temperatures when the magnetic field is aligned perpendicular to (111) plane. (b) Temperature dependence of the corresponding FFT amplitudes, the solid lines represent the fit for effective mass. (c) The Dingle plot of the α and β pockets, the red lines are the fitting result. (d) and (e) The Lifshitz-Kosevich (LK) fit (red line) of the dHvA oscillation pattern of α pocket (yellow circles) and β pocket (blue circles).

To investigate the Fermi surfaces topology of pyrite $PtBi_2$, we performed angular-dependence of dHvA effect measurements with magnetic field rotated from the out of plane to the in-plane direction, as shown in Fig. 4(a). In Fig. 4(b) we depict the oscillation components with the magnetization background are removed. Although the dHvA oscillation patterns vary with field orientations, the oscillations remain to be strong when rotating the field direction, indicating 3D Fermi surface for pyrite $PtBi_2$. From the FFT spectra of the oscillation patterns [Fig. 4(c)], we extracted the angular-dependences of all oscillation frequencies $F_\alpha$, $F_{\alpha'}$, $F_{\alpha''}$, and $F_{\beta}$, as shown in Fig. 4(d). As shown below, these angular-dependences frequencies agree well with the Fermi surface morphology of pyrite PtBi$_2$.

The electronic structure of pyrite PtBi$_2$ was obtained by performing an *ab initio* calculation. The band structure was calculated by density functional theory and fully relaxed atomic coordinates. As shown in the inset of Fig. 4(d), the electronic structure consists of electron (pink) and hole (yellow) Fermi surface in the first Brillouin zone. The twelve ellipsoid-like hole Fermi surface pockets (labeled by 1-12) are unequivalent with the principal axes pointing different directions. From our experimental setup [Fig. 4(b), inset] and angular dependences of frequencies, these hole pockets give rise to the observed oscillation frequencies $F_{\alpha}$, $F_{\alpha'}$ and $F_{\alpha''}$. More specifically, for the strong anisotropy of α pockets in pyrite PtBi$_2$, when magnetic field is parallel to [110] direction (θ=90°), the inequivalence twelve α pockets can be divided into three groups in the first Brillouin zone, i.e., group 1 involving Nos. 1 and 3, group 2 involving Nos. 2 and 4, and group 3 involving Nos. 5-12. Each group contributes a different extremal cross section when magnetic fixed specific direction, resulting in three different oscillation frequencies. The $F_{\alpha'}$ and $F_{\alpha''}$ are result from the oscillations of Fermi pockets in group 1 and group 2 while the main frequency $F_{\alpha}$ is from the oscillation of Fermi pockets in group 3. Therefore, compared with the SdH oscillations that only probed the $F_{\alpha}$ frequency[22], dHvA effect provides opportunities to investigate the complete hole Fermi surface pockets that were not accessible in SdH oscillations. We note that there should be only two frequencies with B//[111] (θ=0°), the observation of three frequencies might be due to the uncertainty about ±5° of the magnetic field orientation in the magnetization measurement. As for $F_{\beta}$, i.e. the light electron (pink) Fermi surface located around the R point of the first Brillouin zone are equivalent, so we detected only one oscillation frequency, which is in accord with the transport results.

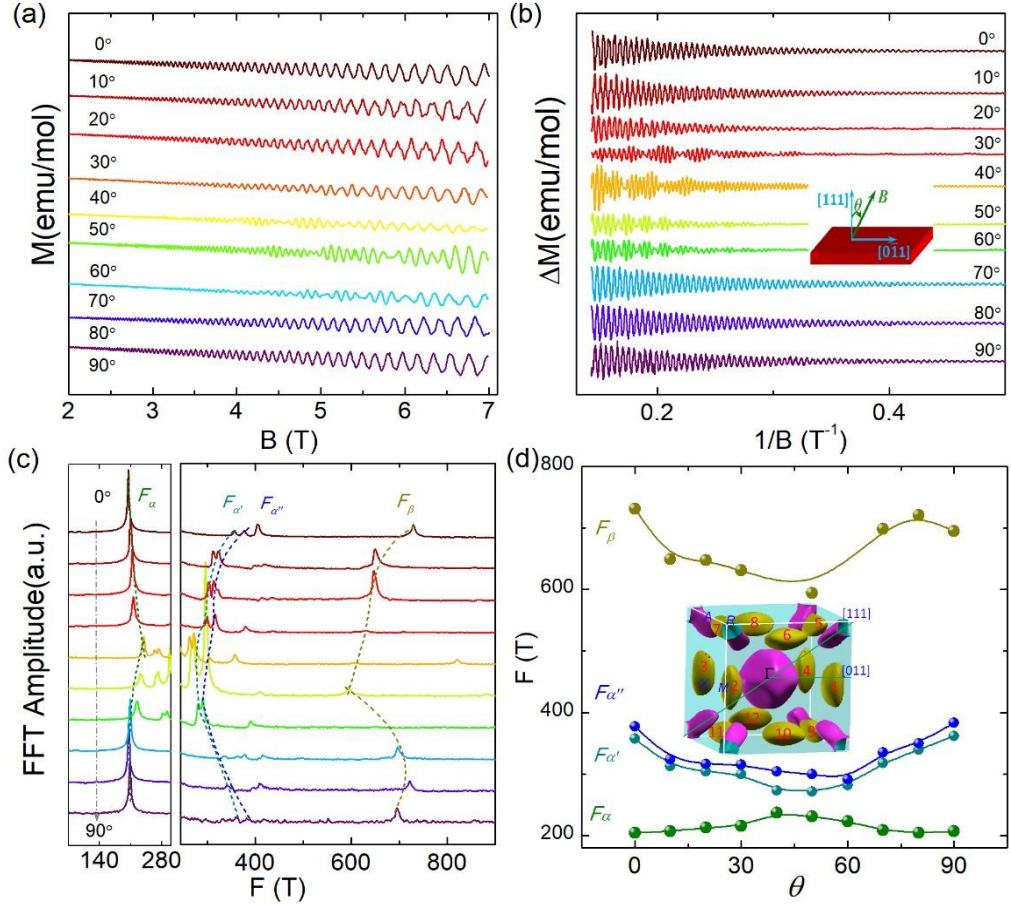

FIG. 4. (a) Magnetic field dependence of magnetization at different angles, $\theta$, at 2K. (b) The dHvA oscillations as a function of 1/B at different angles. (c) The shifted FFT spectra for different angles. Inset: The schematic configuration for angular dependent measurement of magnetization. (d) The angular dependences of the oscillation frequencies of both α and β bands. Inset: The calculated electron (pink) and hole (yellow) Fermi surface of pyrite $PtBi_2$ in the first Brillouin zone.

To make a comparison of the results between the SdH oscillations and dHvA effect, we summarized the major characteristic parameters in TableⅠ. We found that both the main dHvA oscillation frequencies are slight lower than those of the SdH results. The magnetization results also reveal a lighter effective carrier mass and higher quantum mobility than those from the transport results. As we know, the dHvA effect can be well interpreted by LK theory[30], the SdH effect is an analogous effect on the electrical resistance and shares the same physics of the quantized landau level picture with former, but there is delicate difference between them. The dHvA effect is an equilibrium state behavior, the oscillatory magnetization is related to the oscillations of the density of states. The SdH effect is a nonequilibrium dynamics effect, except for the state density, it is also closely related to the electron scattering rate, which has the same oscillation period with state density. However, the electron scattering can be complicated by some mechanisms, such as lattice scattering, impurity scattering and inter/inner-Landau level scattering *et al* [27]. When take them into account, the amplitude of SdH oscillation is only an approximate expression, which will further affect the detection of effective mass. Similar behaviors have been observed in other reported topological semimetals, for examples, in the nodal-line semimetal ZrSiS, different groups obtained different effective masses, and the smallest effective mass from dHvA oscillations is 3 times smaller than the biggest effective mass from SdH oscillations[28]. Besides, in Dirac semimetal LaBi, the effective mass of β band (~600T) obtained from dHvA oscillations[29] is about 2 times smaller than that from SdH oscillations[30]. From this perspective, the original LK theory is more precise to describe the dHvA effect originating from the oscillations of free energy. As a result, the dHvA oscillations will give a lighter effective mass and a smaller dingle temperature of both pockets, resulting in the high quantum mobilities in pyrite $PtBi_2$. The dHvA oscillation also detected additional frequencies compared with the SdH oscillations, which provides a more accurate information of the Fermi surface. Additonally, we note that the effective mass was obtained in the field range of 1.5~7T for dHvA oscillations while 18~33T for the SdH oscillations. The selected different field range between dHvA and SdH oscillations may also affect the effective mass values.

TABLE Ⅰ. Several characteristic parameters derived from the dHvA and SdH oscillations of α and β pockets. *F*, the oscillation frequency; $m_{eff}$, the effective mass; $v_F$, the Fermi vector; $T_D$, the dingle temperature; $\mu_q$, the quantum mobility.

| Bands | α | | | | | β | | | | |
|---|---|---|---|---|---|---|---|---|---|---|
| Quantities | *F* (T) | $m_{eff}$ ($m_0$) | $v_F$ ($10^5$m/s) | $T_D$ (K) | $\mu_q$ (cm$^2$V$^{-1}$s$^{-1}$) | *F* (T) | $m_{eff}$ ($m_0$) | $v_F$ ($10^5$m/s) | $T_D$ (K) | $\mu_q$ (cm$^2$V$^{-1}$s$^{-1}$) |
| dHvA | 204 | 0.17(8) | 5.3(5) | 3.26 | 3940.4 | 730 | 0.43(5) | 4.0(1) | 3.87 | 1299.6 |
| SdH[22] | 250 | 0.64(1) | 1.58 | 7.4 | 453.4 | 850 | 0.68(1) | 2.76 | 10.6 | 298.7 |

In conclusion, we have performed dHvA oscillation studies on Dirac semimetal pyrite $PtBi_2$ using magnetization measurements. Strong dHvA quantum oscillations can be observed at a field as low as 1.5 T. The analyses of the oscillation patterns reveal high quantum mobilities and light effective masses for Dirac fermions in pyrite $PtBi_2$. The angular dependences of the oscillation frequencies agree well with its electronic structure. Furthermore, the observation of additional low frequencies provides us with opportunities to investigate the Fermi surface pockets that were not accessible in SdH oscillations.

This work was supported by the National Key Research and Development Program of China No.2016YFA0401003; the Natural Science Foundation of China (Grant No.11774353, No.11574320, No.11204312, No.11674331, No.11474289, and No. U1432251); the Youth Innovation Promotion Association of Chinese Academy of Sciences (No. 2017483); the Chinese Academy of Sciences Pioneer Hundred Talents Program.

[1] M. Orlita, , D. M. Basko, M. S. Zholudev, F. Teppe, W. Knap, V. I. Gavrilenko, N. N. Mikhailov, S. A. Dvoretskii, P. Neugebauer, C. Faugeras, C. Faugeras, A-L. Barra, G. Martinez and M. Potemski. Nat. Phys.**10**, 233-238 (2014).

[2] X. G. Wan, A. M.Turner, A. Vishwanath, and S. Y. Savrasov. Phys. Rev. B 83, 205101 (2011).

[3] Z. J. Wang, Y. Sun, X. Q. Chen, C. Franchini, G. Xu, H. M. Weng, X. Dai, and Z. Fang. Phys. Rev. B 85, 195320 (2012).

[4] S. M. Young, S. Zaheer, J. C. Y. Teo, C. L. Kane, E. J. Mele, and A. M. Rappe. Phys. Rev. Lett. 108, 140405 (2012).

[5] C. Fang, M. J. Gilbert, X. Dai, and B. A. Bernevig. Phys. Rev. Lett.108, 266802 (2012).

[6] S.Y. Xu, C. Liu, S. K. Kushwaha, R. Sankar, J. W. Krizan, I. Belopolski, M. Neupan, G. Bian, N. Alidoust, T. R. Chang, H. T. Jeng, C. Y. Huang, W. F. Tsai, H. Lin, P. P. Shibayev, F. C. Chou, R. J. Cava, M. Z. Hasan. Science.1256742 (2014).

[7] B. Q. Lv, S. Muff, T. Qian, Z. D. Song, S. M. Nie, N. Xu, P. Richard, C. E. Matt, N. C. Plumb, L. X. Zhao, G. F. Chen, Z. Fang, X. Dai, J. H. Dil, J. Mesot, M. Shi, H. M. Weng, and H. Ding. Phys. Rev. Lett. 115.217601(2015).

[8] Z. K. Liu, J. Jiang, B. Zhou, Z. J. Wang, Y. Zhang, H. M. Weng, D. Prabhakaran, S.-K. Mo, H. Peng, P. Dudin, T. Kim, Hoesch, Z. Fang, X. Dai, Z. X. Shen, D. L. Feng, Z. Hussain, and Y. L. Chen. Nat. Mater.13, 677–681 (2014).

[9] S. Borisenko, Q. Gibson, D. Evtushinsky, V. Zabolotnyy, B. Buchner, and R. J. Cava. Phys. Rev. Lett. 113, 027603 (2014).

[10] Z. K. Liu, B. Zhou, Y. Zhang, Z. J. Wang, H. M. Weng, D. Prabhakaran, S. -K. Mo, Z. X. Shen, Z. Fang, X. Dai, Z. Hussain, and Y. L. Chen. Science 343**,** 864-867 (2014).

[11] S. Y. Xu, C. Liu, S. K. Kushwaha, R. Sanker, J. W. Krizan, I. Belopolski, M. Neupane, G. Bian, N. Alidoust, and T. R. Chang, H,-T, Jeng, C-Y. Huang, W-F. Tsai, H. Lin, P. P. Shibayev, F.-C. Chou, R. J. Cava, M. Z. Hasan. Science 347, 294-298 (2015).

[12] J. Xiong, S. K. Kushwaha, T. Liang, J. W. Krizan, M. Hirschberger, W. Wang, R. J. Cava, N. P. Ong. Science, 350, 413-416 (2015).

[13] L. P. He, X. C. Hong, J. K. Dong, J. Pan, Z. Zhang, J. Zhang, and S. Y. Li. Phys. Rev. Lett. 113, 246402 (2014).

[14] J. Hu, J. Y. Liu, D. Graf, S. M. A. Radmanesh, D. J. Adams, A. Chuang, Y. Wang, I. Chiorescu, J. Wei, L. Spinu and Z. Q. Mao. Sci. Rep. 6, 18674 (2016).

[15] T. Liang, Q. Gibson, M. N. Ali, M. H. Liu, R. J. Cava, and N. P. Ong. Nat. Mater. 14, 280-284 (2015).

[16] W. S. Gao, N. N. Hao, F. W. Zheng, W. Ning, M. Wu, X. D. Zhu, G. L. Zheng, J. L. Zhang, J. W. Lu, H. W. Zhang, C. Y. Xi, J. Y. Yang, H. F. Du, P. Zhang, Y. H. Zhang, and M. L. Tian. Phys. Rev. Lett. 118, 256601 (2017).

[17] A. Pariari, P. Dutta, and P. Mandal. Phys. Rev. B 91, 155139 (2015).

[18] S. Khim, K. Koepernik, D. V. Efremov, J. Klotz, T. Forster, J. Wosnitza, M. I. Sturza, S. Wurmehl, C. Hess, J. Brink, and B. Buchner. Phys. Rev. B 94, 165145 (2016).

[19] J. Hu, Z. J. Tang, J. Y. Liu, Y. L. Zhu, J. Wei and Z. Q. Mao. Phys. Rev. B 96, 045127 (2017).

[20] J. Hu, Z. J. Tang, J. Y. Liu, Y. L. Zhu, D. Graf, K. Myhro, S. Tran, C. N. Lau, J. Wei, and Z. Q. Mao. Phys. Rev. Lett. 117, 016602 (2016).

[21] J. Hu, Y. L. Zhu, D. Graf, Z. J. Tang, J. Y. Liu, and Z. Q. Mao. Phys. Rev. B 95, 205134 (2017).

[22] F. Arnold, C. Shekhar, S. C. Wu, Y. Sun, R. D. Reis, N. Kumar, M. Naumann, M. O. Ajeesh, M. Schmidt, A. G. Grushin, J. H. Bardarson, M. Baenitz, D. Sokolov, H. Borrmann, M. Nicklas, C. Felser, E. Hassinger and B. Yan. Nat. Commun. 7, 11615(2016).

[23] S. Furuseth, K. Selte, A. Kjekshus. Acta Chemica Scandinavica 19, 735-741 (1965).

[24] I. M. Lifshitz and A. M. Kosevich. Sov. Phys. JETP 2, 636 (1956).

[25] G. P. Mikitik and Y.V. Sharlai. Phys. Rev. Lett. 82, 2147 (1999).

[26] J. Hu, J.Y. Liu, D.Graf, S. M.A. Radmanesh, D. J.Adams, A. Chuang, Y.Wang, I. Chiorescu, J.Wei, L. Spinu and Z.Q. Mao. Sci. Rep. 6, 18674 (2016).

[27] E. N. Admas and T. D. Holstein. J. Phys. Chem. Sol. 10, 254 (1959).

[28] M. Matusiak, J. R. Cooper and D. Kaczorowski. Nat. Commun. 8,15219 (2017).

[29] R. Singha, B. Satpati and P. Mandal. Sci. Rep. 7, 6321 (2017).

[30] S. Sun, Q. Wang, P-J. Guo, K. Liu and H. Lei. New J. Phys. 18, 082002 (2016).